# Performance Evaluation of a Natural Language Processing approach applied in White Collar crime investigation

Maarten Banerveld[2], Nhien-An Le-Khac[1] and M-Tahar Kechadi[1]

[1]School of Computer Science & Informatics, University College Dublin
Belfield, Dublin 4, Ireland
`{an.lekhac,tahar.kechadi}@ucd.ie`

[2]School of Computer Science & Informatics, University College Dublin
Belfield, Dublin 4, Ireland
`maarten@imeel.net`

**Abstract.** In today's world we are confronted with increasing amounts of information every day coming from a large variety of sources. People and corporations are producing data on a large scale, and since the rise of the internet, e-mail and social media the amount of produced data has grown exponentially. From a law enforcement perspective we have to deal with these huge amounts of data when a criminal investigation is launched against an individual or company. Relevant questions need to be answered like who committed the crime, who were involved, what happened and on what time, who were communicating and about what? Not only the amount of available data to investigate has increased enormously, but also the complexity of this data has increased. When these communication patterns need to be combined with for instance a seized financial administration or corporate document shares a complex investigation problem arises. Recently, criminal investigators face a huge challenge when evidence of a crime needs to be found in the Big Data environment where they have to deal with large and complex datasets especially in financial and fraud investigations. To tackle this problem, a financial and fraud investigation unit of a European country has developed a new tool named LES that uses Natural Language Processing (NLP) techniques to help criminal investigators handle large amounts of textual information in a more efficient and faster way. In this paper, we present briefly this tool and we focus on the evaluation its performance in terms of the requirements of forensic investigation: speed, smarter and easier for investigators. In order to evaluate this LES tool, we use different performance metrics. We also show experimental results of our evaluation with large and complex datasets from real-world application.

# 1 Introduction

Since the start of the digital information age to the rise of the Internet, the amount of digital data has dramatically increased. Indeed, we are dealing with many challenges when it comes to data. Some data is structured and stored in a traditional relational database, while other data, including documents, customer service records, and even pictures and videos, is unstructured. Organizations also have to consider new sources of data generated by new devices such as sensors. Moreover, there are other new key data sources, such as social media, click-stream data generated from website interactions, etc. The availability and adoption of newer, more powerful mobile devices, coupled with ubiquitous access to global networks will drive the creation of more new sources for data. As a consequence, we are living in the Big Data era. Big Data can be defined as any kind of datasets that has three important characteristics: huge volumes, very high velocity and very wide variety of data. Obviously, handling and analysing large, complex, and velocity data have always offered the greatest challenges as well as benefits for organisations of all sizes. Global competitions, dynamic markets, and rapid development in the information and communication technologies are some of the major challenges in today's industry. Briefly, we have had a deluge of data from not only science fields but also industry, commerce and digital forensics fields. Although the amount of data available to us is constantly increasing, our ability to process it becomes more and more difficult. This is especially true for the criminal investigation today. For instance, a criminal investigation department CID of the Customs Force in a European country has to analyse around 3.5 Terabyte of data (per case) to combat fiscal, financial-economic and commodity fraud safeguards the integrity of the financial system and to combat also organized crime, especially its financial component.

Actually, CID staff focuses on the criminal prosecution of: Fiscal fraud (including VAT/carousel fraud, excise duty fraud or undisclosed foreign assets); Financial-economic fraud (insider trading, bankruptcy fraud, property fraud, money laundering, etc.); Fraud involving specific goods (strategic goods and sanctions, raw materials for drugs, intellectual property, etc.). Seizing the business accounts is usually the first step in the investigation. The fraud must be proved by means of the business accounts (among other things). Investigation officers not only investigate paper accounts, but also digital records such as computer hard disks or information on (corporate) networks. The fraud investigation unit uses special software to investigate these digital records. In this way we gain an insight into the fraud and how it was committed. Interviewing or interrogation of a suspect is an invariable part of the investigation. A suspect can make a statement, but may also refuse this. In any case, the suspect must be given the opportunity to explain the facts of which he is suspected. During their activities the investigation officers can rely on the Information-gathering teams for information and advice. These teams gather, process and distribute relevant information and conduct analyses. With digital investigations we respond to the rapid digitalization of society. This digitalization has led to new fraud patterns and methods, and all kinds of swindle via the Internet. In order to trace these fraudsters we use the same digital possibilities as they do.

As the CID handles around 450 criminal investigations every year, the amount of (digital-) data that is collected increases year over year. A specific point of attention is that the CID operates in another spectrum of investigations as 'regular' police departments. The types of crime that the CID needs to investigate mostly revolve around written facts. So the evidence that is collected by the CID by default contains of large amounts of textual data. One can imagine how much textual data a multinational firm produces, and how many e-mails are being sent in such companies. A specific challenge for law enforcement departments that are involved with fraud investigations is: how can we find the evidence we need in these huge amounts of complex data. Because of the enormity of the large and complex data sets the CID seizes, it is necessary need to look for new techniques that make computers perform some analysis tasks, and ideally assist investigators by finding evidence. Recently, CID has developed a new investigation platform called LES. This tool is based on Natural Language Processing (NLP) techniques [1] such as Named Entity Extraction [2] and Information Retrieval (IR) [3] in combining with a visualization model to improve the analysis of a large and complex dataset.

In this paper, we evaluate the performance of LES tool because there are very few NLP tools that are being exploited to tackle very large and complex datasets in the context of investigation on white-collar crimes. Indeed, theoretical understanding of the techniques that are used is necessary. This theoretical review can help explain the usage of these new techniques in criminal investigations, and pinpoint what work needs to be done before the most effective implementation and usage is possible. The rest of this paper is organised as follows: Section 2 shows the background of this research including related work in this domain. We present briefly LES tool and evaluation methods in Section 3. We apply our method to analysis the performance of LES tool on a distributed platform in Section 4. Finally, we conclude and discuss on future work in Section 5.

## 2   Background

### 2.1   Natural Language Processing in Law Enforcement

NLP implemented techniques can be very useful in a law enforcement environment, especially when unstructured and large amounts of data need to be processed by criminal investigators. Already commonly used techniques like Optical Character Recognition (OCR) [4] and machine translations [5] can be successfully used in criminal investigations. For example OCR is used in fraud investigations to automatically transform unstructured invoices and other financial papers into searchable and aggregated spread sheets. In the past more difficult to implement techniques like automatic summarization of texts, information extraction, entity extraction and relationship extraction [1] are now coming into reach of law enforcement and intelligence agencies. This is manly so because of the decline in cost per processing unit and the fact that these techniques need a large amount of processing power to be able to used effectively.

To zoom in on this a little further: for example the extraction of entities out of large amounts of text can be useful when it is unclear what persons or other entities are involved in a criminal investigation. Combined with a visual representation of the present relations between the extracted entities, this analysis can provide insight in the corresponding (social-) networks between certain entities. Indeed, the usage of NLP techniques to 'predict' criminality, for example grooming by possible paedophiles [6] or trying to determine when hit-and-run crimes may happen by analysing Twitter messages [7] is possible today. A movement from single available NLP techniques like text summarization, text translation, information and relationship extraction towards more intelligent NLP based implementations for law enforcement like crime prediction, crime prevention, criminal intelligence gathering, (social-) network analysis and anomaly detection can be observed in literature. Also theoretical frameworks and models in the field of 'forensic linguistics' [8] are proposed which can be used behind the technical implementation of NLP techniques in criminal investigations.

When (commercial-) solutions using these techniques come available, this could lead to more extensive NLP based law enforcement systems that can handle Crime prediction, deliver automated intelligence on criminal activities, analyse the behaviour of subjects on social networks and detect anomalies in texts or other data. The output of these systems is ideally presented in a visual comprehensible way so the criminal investigator can quickly assess the data and take appropriate action.

## 2.2 Big Data in criminal investigations

No strict definition can be given for the concept Big Data [9] as such, but what can be concluded is that Big Data at least has some common elements and that Big does not necessarily mean large volumes. Complexity and the inner structure of the data are also very important to determine if a dataset belongs to the concept of Big Data or not. Another term that is commonly used when people talk about 'Big Data' is 'Unstructured Data '. As law enforcement we are confronted with at least parts of the Big Data problem; for instance in fraud investigations *the fraud investigation unit* regularly seizes a complete company (network-) environment including cloud storage and all belonging data. Because this data for the *fraud investigation unit* as outsiders is unstructured, and from a variety of sources (computer images, servers, internal communication, wiretap data, databases etc.) these datasets fall under the definition, and elements, of Big Data in terms of volume and complexity (also known as variety of the data). But also a very large e-mail database containing millions of suspect e-mails can fall under the Big Data problem because of the complexity of this data set. Please note that in most descriptions Big Data is measured against three axes: Volume, Variety and Velocity. What we see is that in the *fraud investigation unit's* types of investigation, the focus is mostly on the volume and variety of the large data set. Velocity is not really an issue as they are investigating a static data set. This is so because after seizure the data that needs to be investigated will not (rapidly) change anymore.

What can be said is that the existence of Big Data poses new and unique challenges for law enforcement when evidence needs to be found in an investigation with these characteristics. What also was can be said is that not only the actual size of the total seized data matters, but also the rate of complexity of the data that determines if a case falls under a Big Data definition.

As an example, in a large carousel fraud case that the *fraud investigation unit* investigated in the past, the suspect was a bank that operated from *the fraud investigation unit's* territory and several countries abroad. In this case investigation data was collected and seized from a lot of sources: internet wiretaps, forensic disc images from tens of workstations, user data from server systems, e-mail servers with literally millions of e-mails, company databases, webservers, and the complete banking back-end systems containing all bank transactions. This investigation had the characteristics of Big Data on both levels, a high complexity of the data (the complete banking system had to be reconstructed and analysed) and a high amount of total data (the house searches were in 2006, and in that time a total of 15 terabyte of data was seized).

This paper is about the usage of NLP techniques in fraud investigations, there are specific characteristics for these types of investigations that determine why another approach towards Big Data investigation is necessary. In fact our *fraud investigation unit* mostly investigates White Collar Crime cases. Most police departments focus on other criminal offenses like murder cases, child abuse, threats, hacking, malware etc. The *fraud investigation unit* on the other hand acts on criminal cases like money laundering, terrorism funding, (tax-) fraud, etc. For the *fraud investigation unit* this focus on White Collar crime means that the *fraud investigation unit* has to be able to investigate: (i) Complex (unstructured-) datasets; (ii) Large datasets; (iii) Company networks; (iv) Complex communication networks between suspects; (v) Mostly text based evidence.

As you can see, this list shows that the *fraud investigation unit* will encounter Big Data problems *because* of the specific criminal investigation domain, and that the evidence the *fraud investigation unit* gathers is *mostly* text based. Before the introduction of NLP techniques running on the new *fraud investigation unit* platform LES, the *fraud investigation unit* had massive problems with handling the enormous amounts of data that are so specific for white-collar crime investigations. These problems can be summarized in:

- Time taken to process al data that was seized
- Forensic software not able to handle the huge amounts of data items coming from for instance e-mail databases
- Crashing software when querying the investigation tooling database, because of overload
- Unacceptable waiting time for investigators when performing a query on the data (up to 30 minutes per query)
- Too many search hits to make analysis of the evidence humanly possible in many cases
- Too much technical approach and interfacing for regular investigators by currently used tooling

What also can be observed is that most police cases can make use of the Digital Forensics methodology and tooling as is described in literature [10]. Unfortunately the *fraud investigation unit* has to use tooling that is best suitable for criminal investi-gations falling under Police Types of crime where evidence can be found in/from files, desktop/mobile devices, email, network/memory analysis, etc.

### 2.3 Related work

There are very few researches of NLP in the context of Digital Forensics, especially to tackle the problem of Big Data of financial crimes. In the context of Digital Forensics, [11] used NLP techniques to classify of file fragments. In fact, they use support vector machines [12] along with feature vectors consisted of the unigram and bigram counts of bytes in the fragment. The method proposed is efficient; it is however, not in the context of investigating documents related to financial crimes. In [13], authors proposed a corpus of text message data. This corpus can support NLP techniques in investigating data located on mobile devices. This corpus is very useful in analysing short text but it is not for long, complex documents such as MS word document, presentations, spread sheets, etc. Related to the forensics financial crimes, [14] proposed a semantic search based on text mining and information retrieval. Authors however focus on documents from collaboration platform such as e-mail, forum as well as in social networks. Their main objective is how to optimise the searching queries.

## 3 LES tool and Method of evaluation

In this section, we present briefly LES, a NLP based tool that has been developed to study the possibilities and benefits the usage of NLP techniques can provide in complex fraud investigations. Next, we describe the investigating process where we apply LES tool to analyse evidence files. Finally we present methods we used to evaluate this tool.

### 3.1 LES tool

Because of the problems of handing Big Data investigations mentioned earlier, our fraud investigation unit decided to develop tooling in-house that would be able to handle these specific types of investigations. The three most important requirements for the new tool are:

- Improving the data processing time, to handle large amounts of data
- Improving the data analysis time needed, to handle complex datasets
- Enable end users to perform complex tasks with a very simple interface

This tool was called LES (Figure 1) and its main characteristics are:

- Running on an Apache Hadoop platform [15]
- Ability to handle large amounts of data
- Use NLP techniques to improve evidence finding
- Visualisation of found (possible-) evidence

- A simple web based GUI with advanced search capabilities

In house developed software components allow investigators to rapidly access forensic disk images or copied out single files. MapReduce [16] jobs are then executed over the data to make parallel processing possible over multiple server nodes. Other MapReduce jobs are built in LES tool for text extraction and text indexing. At this moment the following NLP techniques are implemented in LES:

- Information extraction
- Named Entity Recognition (NER)
- Relationship Extraction

The Information and NER extraction process uses a combination of techniques to extract useful information: tabular extraction (for lists of known and described entities), regular expression extraction, and the Stanford NER library also known as CRFClassifier [17]. The relationships between entities are arbitrarily determined by the distance between these entities. If a distance is smaller than a threshold, a relationship between two entities is stored in the LES system (a Hadoop cluster of computers running LES tool). This implementation of relationship extraction is based on co-reference between words, which in system tests appears to perform quite well.

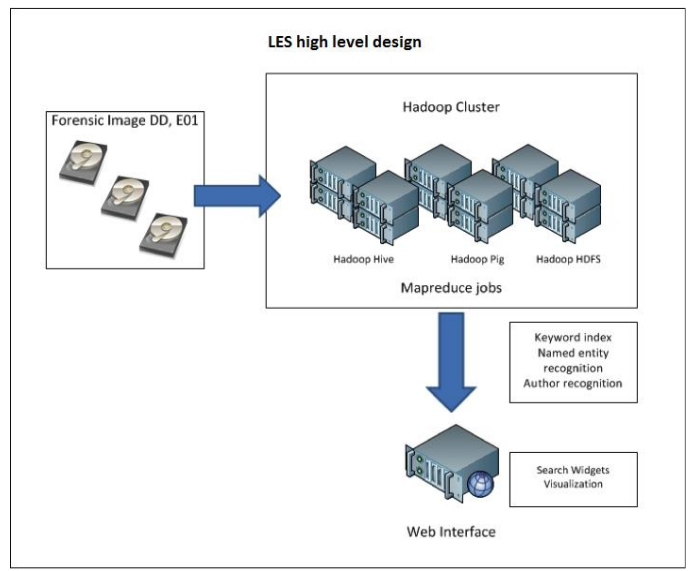

**Fig.1** High level design of LES tool

### 3.2 Investigation Process

Evidence files are imported into the LES system by running specific MapReduce jobs in a predefined sequence:

1) Prepare Evidence

2) Extraction phase
3) Indexing phase
4) NER extraction phase
5) Relationship Extraction
6) Analysing

The evidence acquired during house-searches by the digital investigators is mainly recorded in a forensic format like raw *dd* files or Encase format. During the preparation phase the evidence containers are mounted and integrity is checked. Then by default only the most relevant files are extracted for further processing. At this moment these files are the most common document and textual types and e-mail data. All files that need to be investigated by LES tool are placed in a so-called binary 'blob' or data stream on the Hadoop cluster. Pointers to the original files are recorded in the file index on the cluster. This makes later viewing and retrieval of the original file much easier. When all extracted files are present in the data stream the indexing job is run. Next, the NER and RE phase are performed and finally all results are imported in the LES Hadoop Elastic search environment.

### 3.3 Methodology

Our evaluation method is based on the combination of Microsoft's MSDN performance testing methodology [18], TMap NEXT [19] from Sogeti and some custom evaluation items (quality characteristics). The combination of different methodologies has led to the following concrete test case parameters that were evaluated:

- Test data set processing time, split in time to generate NER, extract relations, generate keyword index
- Test data set query response times
- Accuracy of the data retrieval: cross referencing with standard tooling
- Data controllability: completeness of the data, evidence integrity and chain of evidence reproducibility

## 4    Experiments and Analysis of Results

In this section, we describe firtly the dataset we used in our experiments. We also show the platform where we performed our tests. Finally, we present and analyse the results of these experiments.

### 4.1  Dataset

The dataset that was used is applicable for the two dimensions Volume and Variety (Complexity) of Big Data. Velocity is not an issue for our experiments at this stage. The data set that is used contains historical data from the period 2006 – 2012.

The testing dataset consisted of:

- Total size of dataset: 375GB
- Disk images (Encase E01): 292 disk images
- Microsoft Exchange mail databases: 96GB
- Office documents: 481.000
- E-mails: 1.585.500
- Total size of documents to investigate: 156GB
- Total size of extracted textual data: 21GB

As we are looking for evidence in a fraud case we can expect that most incriminating content can be found in textual data, coming from documents, e-mails etc. LES will automatically extract all files containing textual information out of file containers like Encase images, Exchange databases, zip files etc.

Next, from these files all text is extracted leading to a total size of pure flat text of 21 GB out of a total dataset of 375 GB. As this investigation was performed in the past, we today know that finding and processing the evidence that was needed, took a total of six years investigation. Because the amount of total items to investigate, and the complexity of this dataset, the time needed for this investigation took a lot longer than was thought of at the start. Some statistics can be found as follows:

- Evidence items found in dataset: 2.718
- Total textual items in test dataset: 2.144.254
- Percentage of evidence found versus total textual items: 0,126% (2718 / 2.144.254) x 100 = 0,126%)

As we can see, the percentage of usable evidence for this case was only 0,126 percent. This indicates the needle in the haystack problem we are facing for these types of investigations.

### 4.2 Testing Platform

The testing system is a cluster consists of 14 physical servers with the following roles:
  - 2x Hadoop Namenode (for redundancy purposes)
  - 6x Hadoop Datanode (to store the data on)
  - 1x Hadoop Edgenode (for cluster management)
  - 4x Index nodes (to process the data)
  - 1x webserver (for the end-user GUI)

Hadoop processing and storage:
- 18TB storage
- 24 cores Intel Xeon E5504

Index nodes processing and storage:
- 12TB storage
- 12 cores Intel Xeon E5504

Total cluster internal memory is 256GB. The cluster has been build and configured according to the Hadoop recommendations for building a Hadoop cluster.

### 4.3  Result Description and Analysis

We evaluate first of all the performance perspective of LES tool. We compare the processing time between Forensic Toolkit (FTK) [20] and LES tool on the same testing dataset. FTK has been configured in such a way that it approached the LES way of processing data the most. That means that all images and container items were read in FTK, but all extra options were disabled to make comparison fairer (Figure 2). As you can see, FTK has been configured to not perform Entropy test, carving, OCR. Indeed, only possible textual files were added as evidence to the case (documents, spreadsheets, e-mail messages). Only from this selection FTK was allowed to make a keyword based index.

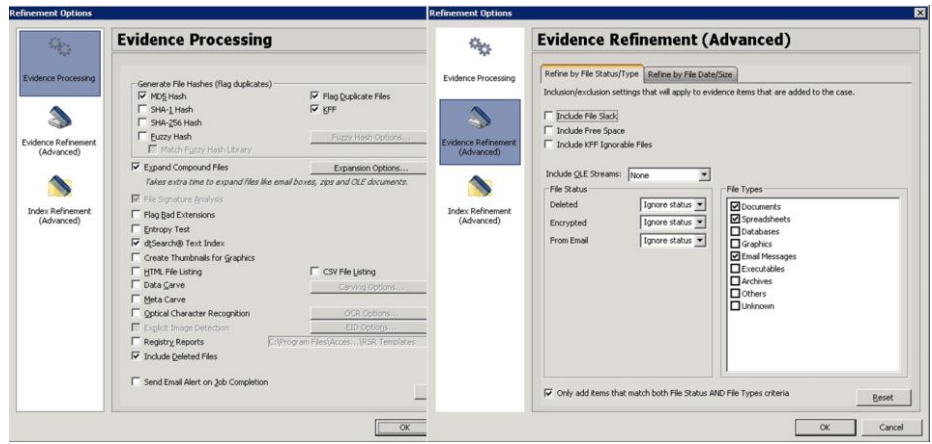

**Fig.2** FTK case configuration

According to the FTK processing log, the FTK processing time of the testing dataset, with the criteria shown above is 10 hours and 54 minutes. For LES tool, the total processing time is 1 hour and 24 minutes including 34 minutes of text extraction, 38 minutes of generating NLP NER databases and 12 minutes of generating the searchable keyword index based on the testing datasets.

In fact, the LES tool was evaluated by running various experiments on the testing datasets. As we can see, the overall processing time of this tool is 6 times faster than FTK with the same testing datasets. Furthermore, when LES is configured to only create a keyword based index, similar to Forensic Toolkit, the LES tool is even 11 times faster than FTK running on the same datasets. LES does however need extra processing time to perform the needed NLP calculations, but enhances the ease of finding evidence by running these NLP processes.

Besides, response times are significantly better when LES is used to search through the test data by using single keywords. Table 1 shows the response time of keyword searching. FTK shows some remarkable slow response times when using single keywords to search for, whereas LES in most cases is a factor 1000 or faster when searching through the data.

**Table 1.** Response time of single keyword search

| Document | Keywords | FTK | | LES | |
| --- | --- | --- | --- | --- | --- |
| | | *Response Time(s)* | *Retrieve Time(s)* | *Response Time(s)* | *Retrieve Time(s)* |
| *D-195* | discontinue | 6 | 27 | 0.005 | < 0.5 |
| *(.msg)* | fraud | 3 | 41 | 0.078 | < 0.5 |
| | revenue | 6 | 287 | 0.119 | < 0.5 |
| *D-550* | scrubbing | 6 | 17 | 0.081 | < 0.5 |
| *(.ppt)* | blacklists | 5 | 14 | 0.099 | < 0.5 |
| | violation | 6 | 365 | 0.069 | < 0.5 |
| *D-718* | [NAME1] | 18 | 295 | 0.138 | < 0.5 |
| *(.xls)* | training | 6 | 383 | 0.143 | < 0.5 |
| | crimecontrol | 7 | 4 | 0.195 | < 0.5 |
| *D-735* | [NAME2] | 3 | 22 | 0.006 | < 0.5 |
| *(.msg)* | [NAME3] | 5 | 52 | 0.023 | < 0.5 |
| | NewYork | 5 | 195 | 0.141 | < 0.5 |
| *D-805* | [NAME4] | 5 | 4 | 0.038 | < 0.5 |
| *(.txt)* | Bermuda | 5 | 1190 | 0.091 | < 0.5 |
| | [NAME5] | 5 | 33 | 0.020 | < 0.5 |

Furthermore, the retrieval times in LES tool are always under 0.5 seconds, but cannot be shown in a counter, and therefore difficult to give an exact count in milliseconds.

However, this is not a very efficient approach to find unique evidence items, most of the time using only one keyword leads to long lists of results. Also, it is very difficult to make up the best fitting keyword to find the evidence. Normally a combination of keywords or multiple search iterations is used, but in practice our investigators start hypothesis building by trying single keywords and see what comes back as a result. What can be seen from the FTK evaluation is that when using single keywords when trying to pinpoint evidence, the retrieval time can be very long. When using single keywords only and keywords are not chosen well or are not unique enough, waiting time becomes unacceptable long from an end-user perspective when we choose a response time of 20 seconds maximum. Of course investigators also need to be trained to perform smart search actions when using FTK. It is essential to choose keyword combinations well. Next, we use the five known evidence items to locate and use an AND combination of keywords to evaluate response time and retrieval time of evidence items. Table 2 shows the results of this experiment.

**Table 2.** Response time of combined keyword search

| Document | Keywords | FTK | | LES | |
| --- | --- | --- | --- | --- | --- |
| | | Response Time(s) | Retrieve Time(s) | Response Time(s) | Retrieve Time(s) |
| D-195 (.msg) | discontinue, fraud, revenue | 28 | 7.1 | 0.006 | < 0.5 |
| D-550 (.ppt) | scrubbing, blacklists, violation | 24 | 5.2 | 0.138 | < 0.5 |
| D-718 (.xls) | [NAME1], training crimecontrol | 10 | 18 | 0.195 | < 0.5 |
| D-735 (.msg) | [NAME2], [NAME3], NewYork | 11 | 5 | 0.232 | < 0.5 |
| D-805 (.txt) | [NAME4], Bermuda, [NAME5] | 16 | 4 | 0.004 | < 0.5 |

When a combination of keywords is used we see that the response times for FTK are worser than single keyword search. On the other hand, using multiple keywords in LES the response time is also milliseconds for the performed evaluations. From an end-user perspective a response in milliseconds is more or less instant and thus leading to a better investigation experience. When the various NLP techniques are used to search for evidence, the LES tool response times are also in milliseconds. For instance the selection of named entities and the drawing of a relation diagram are performed very fast by the system.

Next, we evaluate the total amount of processed evidence items per file type analysing system and processing log files for FTK and LES (Table 3).

**Table 3.** Total number of processed evidence items per file type

| Document Type | Number of items per file type | |
| --- | --- | --- |
| | FTK | LES |
| Email | 1.585.500 | 1.641.063 |
| Word documents | 44.105 | 44.837 |
| Spreadsheets | 68.101 | 38.580 |
| Presentations | 6.548 | 2620 |

As we are not sure how FTK counts evidence items, and which types are counted and which are not, it is difficult to draw a conclusion from these figures. But what we do know is that FTK counts for instance every OLE object item as a unique evidence item for Microsoft Office documents. So that increases the count for FTK significant-ly. However, since we have found all our randomly selected evidence items in both FTK and LES we can be carefully positive that no essential data is lost in LES.

Looking at the functionality perspective, LES has more possible search paths towards an evidence item; this could mean that evidence can be found faster using LES, because an investigator has more chance 'hitting' a useful search path. This coincides with the fact that in LES evidence can be found in more ways, because more search methods are implemented. These search methods increase the ways investigators can search for evidence. Especially the implemented NLP entity selection in combination with other search methods creates new evidence finding possibilities that previously were not possible. When looking at the data presentation parts of the software evaluation we can see that LES has more ways of presenting data to the investigator; the visualization view of found evidence can help investigators finding new leads. Another important functional part is the integrity and chain of evidence of the data. What can be seen is that FTK has better data control embedded, thus in FTK the chain of evidence is maintained more thoroughly. Also, FTK has better file control embedded; tracing back a file to its originating location is better implemented in FTK than in LES, thus the chain of evidence is maintained better. A big advantage of LES is that LES has been developed with the end-user in mind, in this case a financial and fraud investigator who needs to investigate a Big Data set. Specifically the LES query interface is very flexible and helps analyzing complex and large data sets, especially the possibility to add query windows (widgets) and refine searches by doing that is very powerful.

Specific evaluation requirements that were mostly focused on the implementation and usage of NLP techniques show that the implemented NLP techniques can help investigators finding evidence in another way, possibly faster and more efficient. At the minimum a new view towards complex data is presented for investigators. LES requires less search iterations to find evidence, because of the implementation of NLP NER and visualisation of evidence. On the other hand, FTK's keyword based search requires investigators to work through more data and refine search queries a lot of times.

Indeed, some noteworthy points that also came up during the evaluation were for instance that it was difficult to find literature that evaluates Accessdata forensic toolkit on a performance and data controllability level. It looks like this tooling has not been evaluated very thoroughly yet by a respectable authority. For Hadoop/MapReduce techniques we found that the usage of a Hadoop cluster seems to be very efficient when one needs to process large amounts of textual data. However, the programming paradigm of Hadoop/MapReduce are more complex than regular programming problems because of the distributed and multi-processing nature of the Hadoop cluster. The issues that we found during the evaluation were that a (too-) large edges and nodes file leads to graphical representation problems. Too much named entities and extracted relations leads to information overload for the end-user. The forensic chain of evidence is more difficult to maintain in LES. This is because of the nature of LES' inner workings, and the fact that it extracts textual information out of forensic images.

At organization level, we found that the CID will need to explain the difference between forensic computer investigation and analysis of Big Data. When to use what tool all depends on the type of investigation, the needed evidence, and the amount and complexity of the data. As the CID mainly has large fraud cases, a logical choice would be to use LES as the preferred tool for these kinds of investigations. One remark that must be made is that all data found in LES must be verified using a (forensic-) tool until LES has a proven track record in court of law.

As a conclusion, the usage of LES tool that uses NLP as key enabler to handle very large and complex data investigations. This means LES tool improves the 'white collar crime' investigation process in terms of speed and efficiency.

## 5   Conclusions and future work

In this paper, we present and evaluate LES tool that is based on NLP techniques to help criminal investigators handle large amounts of textual information. In fact, we evaluate different perspectives of LES tools. In terms of speed: the proposed solution is significantly faster in handling complex (textual) data sets in less time compared to traditional forensics approach. In terms of efficiency: the proposed solution is optimized for the fraud investigation process. The usage of NLP techniques helps in optimizing the investigation process. Investigators have more possibilities finding evidence in very large and complex dataset, aided by smart NLP based techniques. This greatly improves fraud investigation efficiency.

Some topics for further scientific and practical research is coming up. In terms of LES tool, more functions have being added such as automatic summarization of texts, author recognition, detection of cover language, detection of communication patterns, language detection, adding fraud domain knowledge to a NLP language corpus, visualisation of searching results, etc. Therefore, we will also evaluate the performance of these upcoming features.